\documentclass[a4paper,fleqn,usenatbib]{mnras}

\usepackage{newtxtext,newtxmath}

\usepackage[T1]{fontenc}
\usepackage{ae,aecompl}

\usepackage{graphicx}	
\usepackage{amsmath}	
\usepackage{amssymb}	

\usepackage[flushleft]{threeparttable}

\newcommand{\so}{4U1700$-$37 }
\newcommand{\chandra}{{\it Chandra }}
\newcommand{\hst}{{\it HST }}
\newcommand{\rxte}{{\it RXTE }}

\newcommand{\xmm}{{\it XMM-Newton }}
\newcommand{\ka}{K$\alpha$}

\title[Compton cooling in 4U1700$-$37]{Evidence of Compton cooling during
  an X-ray flare supports a neutron star nature of the compact object in 
4U1700$-$37}

\author[M. Martinez-Chicharro et al.]{
M. Martinez-Chicharro$^{1}$\thanks{E-mail: maria.chicharro@ua.es}
J. M. Torrej\'on$^{1}$
L. Oskinova$^{2}$
F. F\"urst$^{3}$
K. Postnov$^{4}$
\newauthor J.J. Rodes-Roca$^{1}$
R. Hainich$^{2}$
and A. Bodaghee$^{5}$
\\
$^{1}$Instituto Universitario de F\'isica Aplicada a las Ciencias y las
Tecnlog\'ias, Unversidad de Alicante, 03690 Alicante, Spain\\
$^{2}$Institute for Physiscs and Astronomy, Universit\"at Potsdam,
14476 Potsdam, Germany\\
$^{3}$European Space Astronomy Centre (ESAC), Science Operations
Departement, 28692 Villanueva de la Ca\~nada, Madrid, Spain\\
$^{4}$Sternberg Astronomical Institute, Moscow M.V. Lomonosov State
University, 119234 Moscow, Russia\\
$^{5}$Dept. of Chemistry, Physics and Astronomy, Georgia College, 221 N. Wilkinson St., Milledgeville, GA 31061, USA
}

\date{Accepted XXX. Received YYY; in original form ZZZ}

\pubyear{2017}

\begin{document}
\label{firstpage}
\pagerange{\pageref{firstpage}--\pageref{lastpage}}
\maketitle

\begin{abstract}
Based on new {\em Chandra} X-ray telescope data, 
we present empirical evidence of plasma Compton cooling during a flare
in the non pulsating massive X-ray binary 4U1700$-$37. This behaviour
might be explained by quasispherical accretion onto a slowly 
rotating magnetised neutron star. In quiescence, the neutron star in \so is
surrounded by a hot radiatively cooling shell. Its presence is supported by the detection of mHz quasi periodic oscillations likely 
produced by its convection cells. The high plasma temperature 
and the relatively low X-ray luminosity observed during the
quiescence,  point to a  
small emitting area $\sim 1$\,km, compatible with
a hot spot on a NS surface. The sudden transition from a radiative to a significantly
more efficient Compton cooling regime triggers an episode of enhanced 
accretion resulting in a flare. During the flare, the plasma
temperature drops quickly.  The predicted luminosity for such
transitions, $\sim 3\times 10^{35}$ erg s$^{-1}$, is very close to the 
luminosity of \so during quiescence.  The transition may be caused by 
the accretion of a clump in the stellar wind of the donor star. Thus,
a magnetised NS nature of the compact object is strongly favoured.
\end{abstract}




\section{Introduction}

High Mass X-ray Binaries (HMXBs) are fundamental laboratories where
the structure of the stellar wind in massive stars as well as the
physics of accretion onto compact objects can be studied in detail
\citep[for a recent review]{2017SSRv..tmp...13M}. 
One of the best studied HMXBs in the Galaxy, 4U1700$-$37,  
consists of a O6Iafcp donor star (V$^{*}$V884 Sco; \cite{2014ApJS..211...10S}) 
and a compact object on the 3.41\,d orbit 
at an average orbital distance of $a\approx 1.6 R_{*}$. 

Despite the vast amount of multiwavelength observational data accrued so far, 
the nature of the compact object in 4U1700$-$37, remains a  mystery. No coherent 
pulsations have ever been found in X-rays or any other wavelengths. 
The mass determinations of the compact object give $2.44\pm 0.27 M_\odot$ 
\citep{2002A&A...392..909C}, quite high for a neutron star\footnote{although 
still below the
  Tolman-Oppenheimer-Volkoff limit of $\sim 3$ M$_{\odot}$} (NS), but too low
compared with the smallest black hole (BH) found so far. Indirect evidence
supporting the presence of a NS has been provided 
based on the X-ray spectra \citep{2016ApJ...821...23S} or
the X-ray colour-colour behaviour \citep{2013MNRAS.428.3693V}. A BH, in
turn, has been favoured based on timing properties \citep{2011arXiv1107.1537D}.

The donor star in \so, V$^{*}$V884 Sco, is one of the most massive stars 
($M_{*}=58\pm 11M_{\odot}$) known in the Galaxy and the most massive donor  
known in any Galactic HMXB. The inner parts of OB-star winds ($a< 2R_{*}$)
are inhomogeneous and clumped \citep{2015ApJ...810..102T} but their  
properties are not well known. 
In the past, several studies have used the compact object in \so to probe in 
situ the donor star wind. 
\citet{1989ApJ...343..409H}, using
{\em EXOSAT}, studied the radial wind density stratification, via photoelectric
absorption. \citet{2005A&A...432..999V} performed a study of the
emission lines excited in the stellar wind by the powerful X-ray
source, using {\em XMM-Newton}. Hints of highly
ionized iron were detected but could not be completely disentangled from
the nearly neutral Fe \ka\ line at the {\it EPIC-CCD} resolution. The \xmm 
light curves showed an {\it off state}, when the X-ray count 
rate dropped  to zero for a short time. \citet{2005A&A...432..999V} 
attributed this off state to the gap between two successive clumps 
propagating in the stellar wind. 


In this {\it Letter} we present a 14 ks \chandra\ observation of
\so during which the source flared. This enables us to investigate 
the changes in spectral properties at different luminosities. 

The paper is structured as follows: in Section \ref{sec:obs} we
present the observational details. In sections \ref{sec:lc} and
\ref{sec:spec} we analyse the light curve and time resolved spectra of
the source, providing the best fit parameters for the continuum and
the Fe emission lines. Finally, in Sections \ref{sec:disc} and
\ref{sec:conc} we discuss these parameters in the framework of the
theory and present the conclusions.

\section{Observations}
\label{sec:obs}

\begin{table}
	\centering
	\caption{Observations journal}
\begin{threeparttable}
	\label{tab:obs}
	\begin{tabular}{lccr} 
		\hline
		ObsID & Date & $t_{\rm exp}$ & $\phi _{\rm orb}$ \\
		\hline
		17630 &  2015-02-22 03:11:16 & 14.27 &
                0.13\tnote{a}\\
		\hline
	\end{tabular}
\begin{tablenotes}
\item[a] Mid eclipse time $T_{0}=49149.412\pm 0.006$ MJD, orbital period
$P=3.411 660\pm0.000 004$ d \citep{2016MNRAS.461..816I}
\end{tablenotes}
\end{threeparttable}
\end{table}

The DDT observations of \so were performed  by \chandra on 22 
Feb. 2015, simultaneously with the {\em Hubble Space Telescope} (\hst,
P.I. L. Oskinova)\footnote{The \hst\ UV observations will be reported in the follow up publication, 
Hainich et al. in prep.}. The High Energy Transmission Gratings 
spectrometer on-board of \chandra\ (\textsc{hetg}; \citet{2005PASP..117.1144C}) 
acquired data during a total of 14.27 ks. The HETG provides spectra with 
two sets of gratings, the High Energy Grating (\textsc{heg}) which offers a 
resolution of 0.011 \AA\ in the bandpass of about 1.5 to 16 \AA,
and the Medium Energy Grating (\textsc{meg}) which offers a resolution 
of 0.021 \AA\ in the the bandpass of about 1.8 to 31 \AA.
The spectra 
were reduced using standard procedures with the \textsc{ciao} software (v
4.4) and the response files were generated (\texttt{arf} and 
\texttt{rmf}). First dispersion orders ($m=\pm 1$) were extracted and analysed 
simultaneously. The peak source flux both at \textsc{heg} and
\textsc{meg} gratings is 4.4\,s$^{-1}$, which is  much lower than the
level at which pileup starts to be important\footnote {See \emph{The
Chandra ABC Guide to Pileup}, v.2.2, 
\texttt{http://cxc.harvard.edu/ciao/download/doc/pileup-abc.pdf}}. The
spectral analysis was performed with the Interactive Spectral Interpretation
System (\textsc{isis}) v 1.6.1-24 \citep{2000ASPC..216..591H}.

\section{Light curves and timing}
\label{sec:lc} 

\begin{figure*}
\includegraphics[angle=0,width=1.07\columnwidth]{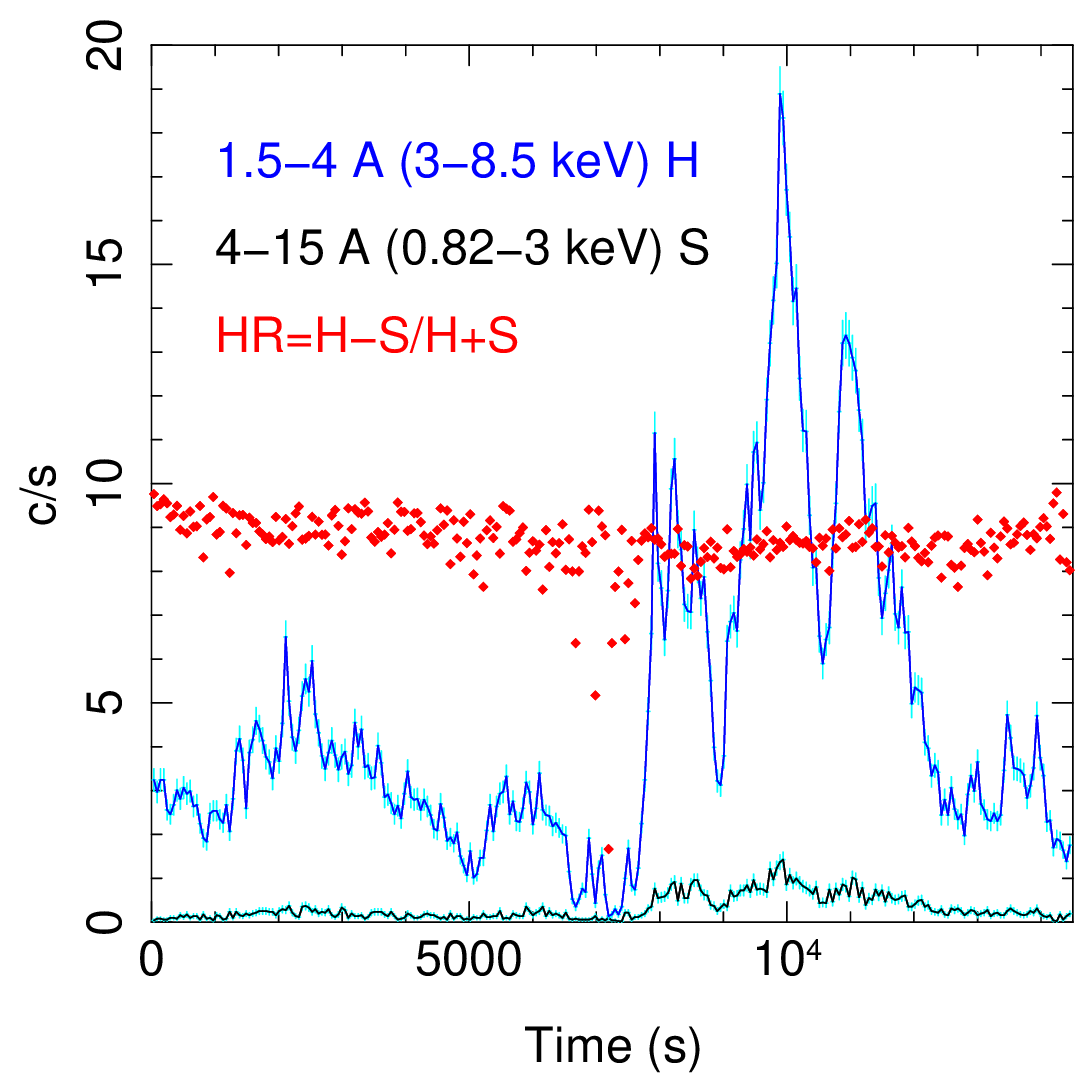}
\includegraphics[angle=0,width=1\columnwidth]{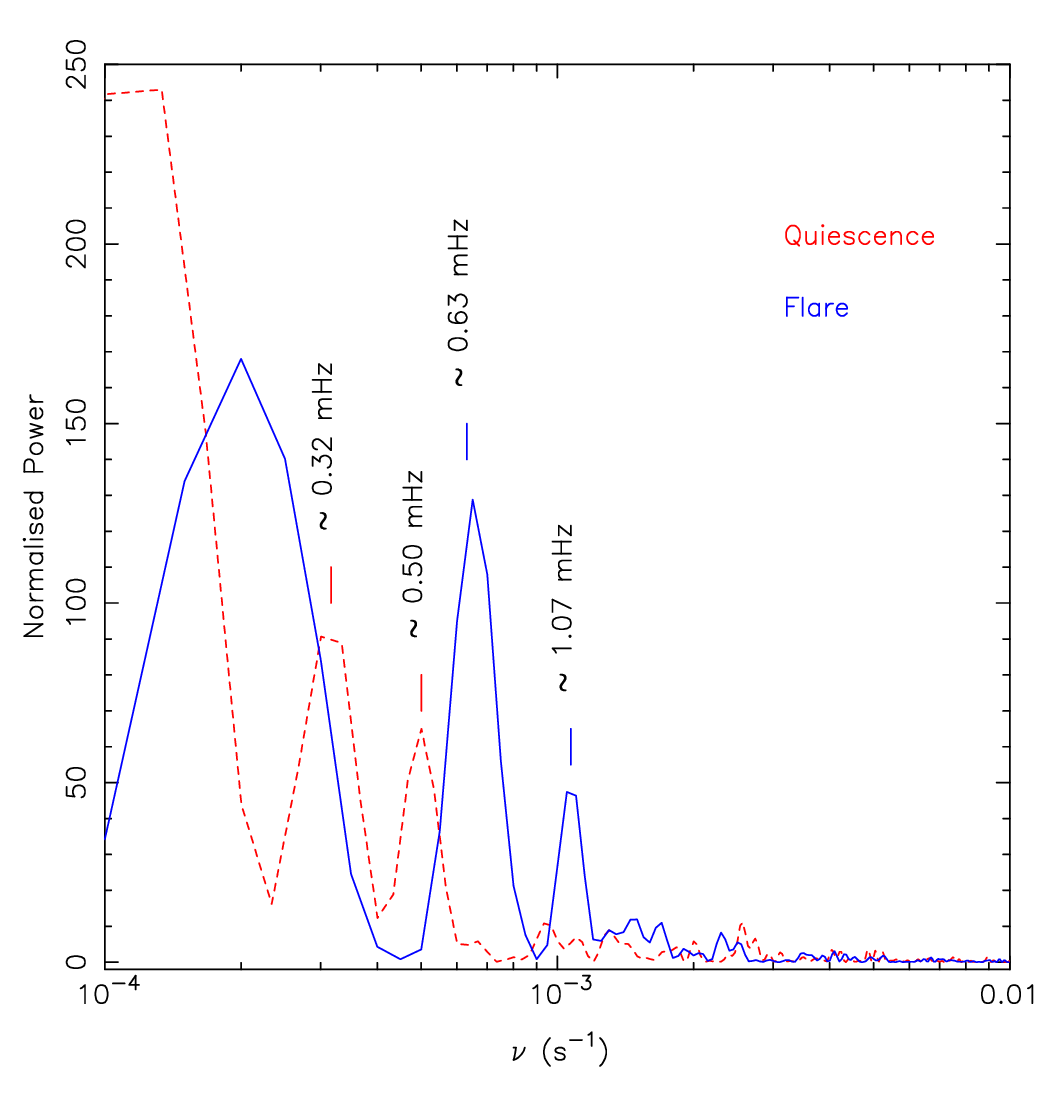}
\caption{\chandra HETG lightcurve, in the 1.2 - 15 \AA\ wavelength
  range, grouped in 10 s bins. The HR seems to be constant even during
  the flare. Right panel: Lomb-Scargle periodogram of the \chandra HETG lightcurve
  binned to 5s for quiescence and flare. No
  coherent period is found but a number of $\sim$ mHz QPOs are
  detected, most prominently during the flare. }
\label {fig:chandra_lc}
\end{figure*}

Figure \ref{fig:chandra_lc} shows the extracted light curves in
the hard 1.5 - 4\,\AA\ (H) and soft 4 - 15\,\AA\ (S) bands. 
Both light curves are strongly variable as typical in wind accreting HMXBs. 
Two episodes are remarkable. The most obvious one lasts from 
$\sim 7600$\, s to 13000\,s when the source undergoes a flare after 
which it returns to the quiescence brightness levels. The hardness 
ratio, HR=H-S/H+S, does not show any noticeable change, even during the flare. 
The second episode is the period of very-low flux, between 7200\,s and 7350\,s, 
preceding the flare. During the low flux period, the count-rate in the soft 
band is consistent with zero counts. Similar {\it off-states} are often
observed in systems containing magnetised NS (eg.\, Vela X-1;
\citet{2008A&A...492..511K}). To our knowledge, the off-states
have not been observed in HMXBs that contain a black hole.

In the right panel of Fig. \ref {fig:chandra_lc} we show the 
standard Lomb-Scargle periodogram of \chandra \textsc{hetg} 1.5-10 \AA\ 
light curve, binned to 5\,s, separately for the quiescence and flare. 
In agreement with previous observations, we do not see any signatures  
of coherent pulsations. However, the periodogram  clearly shows several 
mHz quasi periodic oscillations (QPOs) more noticeable during the flare. 
The false alarm probability of the $0.63\pm 0.05$ 
mHz peak is lower than 1\%. Such QPOs at
mHz frequencies have already been reported for \so\ on the basis of 
\chandra\  \citep{2003ApJ...592..516B} and \rxte\ observations
\citep{2011arXiv1107.1537D}.

\section{Spectra}
\label{sec:spec}

\subsection{Continuum modeling}
\begin{figure}
\includegraphics[angle=0,width=1.0\columnwidth]{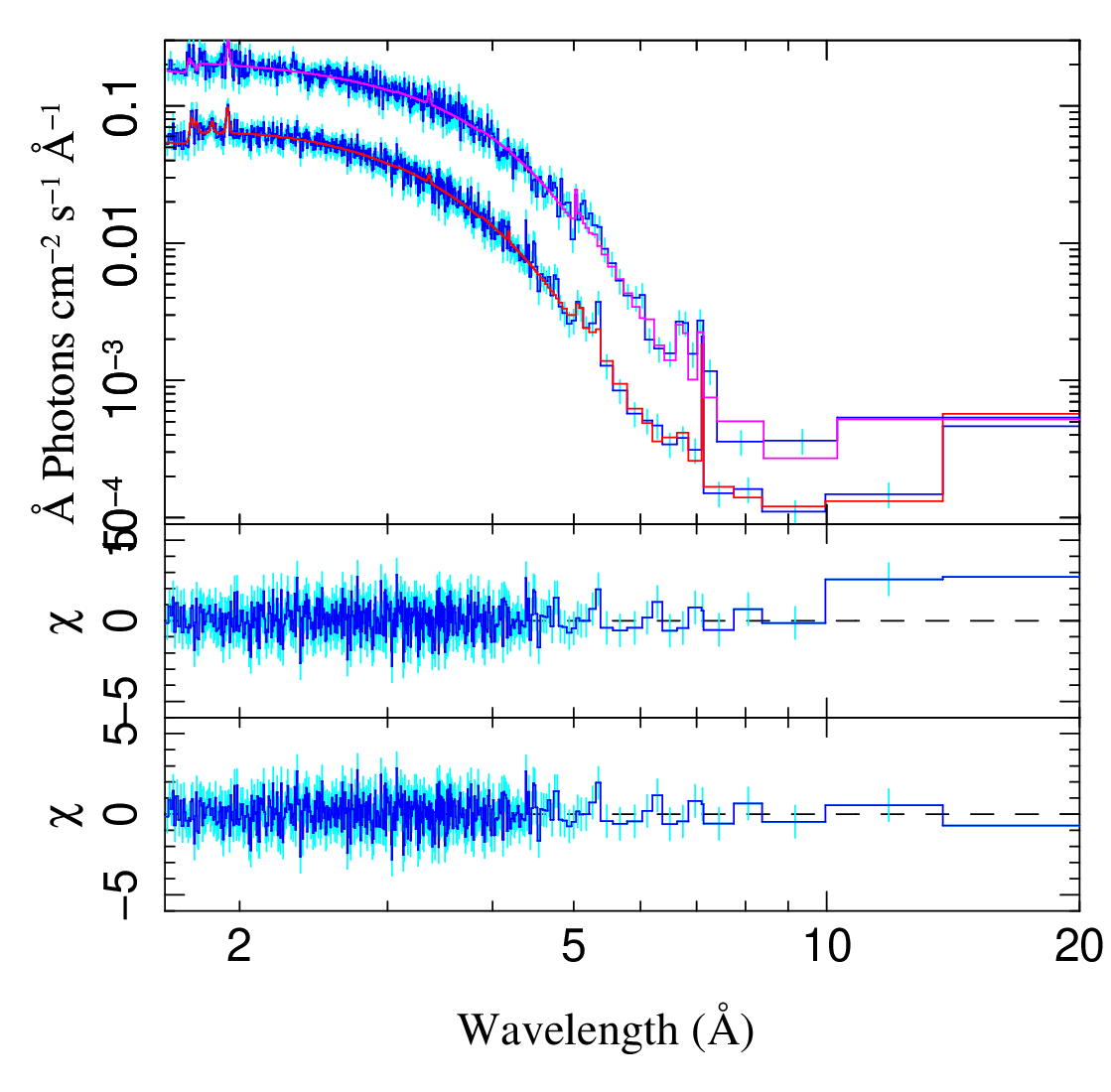}
\caption{\chandra HETG spectra during quiescence and flare grouped in
  5$\sigma$ bins. Residuals
  are shown for the \textsc{bmc} model (mid panel) during quiescence
  (residuals during flare are similar).  The {\it soft excess} seen at
  long wavelengths, found in the spectra of many HMXBs
\citep{2004ApJ...614..881H},  can be described by the addition of a blackbody
  with $N_{\rm H,3}$ (10$^{22}$
  cm$^{-2}$)$=0.3\pm 0.1$ and $kT_{\rm
    bb}=0.095^{+0.007}_{-0.011}$ keV (bottom panel). This component, however, is
  marginal and its addition
  here does not improve the fit.} 
\label {fig:f+q_spec}
\end{figure}

In order to shed light on the mechanism that triggered the flare, we
want to trace plasma changes by performing time resolved spectroscopy during 
the time intervals $t=[0-7600, 13000-14200]$\,s (quiescence) and 
$t=[7600-13000]$\,s (flare). We successfully fit the continuum with the bulk 
motion
comptonisation model (\textsc{bmc})
\citep{1997ApJ...487..834T}.  \textsc{bmc} is a general 
model 
for comptonisation of soft photons
which uses the Green's (spread) functions for the treatment of
upscattering and which attains the form of a broken powerlaw. This
formalism is valid for any kind of comptonisation (bulk comptonisation
in first order $(v/c)$, thermal comptonisation in second order
$(v/c)^{2}$) and remains valid up to photon energies comparable to the
mean plasma energy ($m_{\rm e}c^2\sim 511~\rm keV$ in the case of
bulk motion). In the case of a BH, the soft component originates in
the innermost part of the 
accretion disk where the gravitational energy of matter is released
due to viscous dissipation and geometric compression. In the
case of a NS the soft component is likely produced by a hot spot on the NS 
surface. 
Either the disk or the surface (or both) emit a soft black body
like spectrum with a characteristic colour temperature  $kT_{\rm
  col}$. The comptonizing region (a cloud or a boundary layer)
must cover effectively this zone (i.e. the innermost region of the
disk or the spot over the surface) in order to be well exposed to a high
fraction of the seed photons. Parameter $f$ describes the ratio of the 
number of photons multiply-scattered in the converging inflow to 
the number of photons in the thermal component. During fitting procedure 
we have fixed it at 10 ($\gg 1$) meaning that the Compton cloud efficiently 
covers the soft photons
source\footnote{Given the limited energy range covered by \chandra, the model is not sensitive to the exact value of $f$}. 

One of the three free parameters of the \textsc{bmc} model is a power law
spectral index $\alpha$ describing the Comptonisation efficiency.
 When $\alpha$ is smaller the spectrum is harder due to 
enhanced efficiency  (for details see \citet{1980A&A....86..121S}). On the 
contrary, larger $\alpha$ describes softer spectra. A value close to unity 
indicates that the source is undergoing a phase transition from the low-hard to 
a high-soft state. The transitions can be caused by a number of mechanisms, 
e.g.\ the redistribution of mass accretion rates between Keplerian disk and 
sub-Keplerian components in a disk or by an increase of the optical depth 
$\tau_0$ for gravitational energy release at the shock (BH case) or at the 
surface (NS case). 

The emitted X-ray continuum described above must be modified at low
energies by photoelectric absorption {\sc abs(E)} in order to
account for the local and interstellar absorption effects. The
observed continuum is thus
$F(E)=${\sc Abs(E)}$ \times$ {\sc bmc(E)} where the absorber is described by 

\begin{equation}
{\sc Abs(E)}=\left (\epsilon \times {\rm e}^{-\sigma(E) N_{\rm H,1}} + 
(1-\epsilon)\times {\rm e}^{-\sigma(E)
    N_{\rm H,2}}\right )
\end{equation}

Here $\epsilon$ is the {\it covering fraction} which acts as a proxy for the
massive star wind clumping. The photoelectric absorption has been modelled using
\texttt{tbnew} which contains the most up to date cross sections for
X-ray absorption\footnote{
http://pulsar.sternwarte.uni-erlangen.de/wilms/research/tbabs\\/index.html}. The best fit parameters are presented in
Table \ref{tab:continuum_bcm}.


During quiescence, the column density 
$N_{\rm H,2}=N^{\rm ISM}_{\rm H}=0.30^{+0.28}_{-0.23}\times 10^{22}$
cm$^{-2}$ is
compatible with the interstellar medium (ISM) absorption towards V$^{*}$V884 Sco,
deduced from the optical extinction, $E(B-V)=0.54\pm 0.02$
\citep{2002A&A...392..909C} and using the relationship $N_{\rm H}=6.12\times
10^{21} E(B-V)$ \citep{2012ApJS..199....8G}. On the other hand,
$N_{\rm H,1}$ is much
larger during the flare, consistent with the thick stellar wind expected in a O6.5
supergiant star. 

The spectral powerlaw index $\alpha$ goes from $\sim 1.2$
during quiescence to $\sim 0.19$ during the flare showing that the
comptonisation is more efficient when the luminosity (presumably the
mass accretion rate) increases. At the
same time, the soft photon source temperature $kT_{\rm
  col}$ decreases from 1.43 keV ($\sim 16$ MK) to 0.58 keV (7 MK) during the
flare when the comptonisation (hence the Compton cooling) is more
efficient. Thus we observe plasma that cooled down by about 9 
million degrees on the time scale of an hour. 

\begin{table}
{\def\arraystretch{1.3}
\begin{center}
\caption{Model \textsc{bmc} continuum parameters.}
\begin{threeparttable}
\label{tab:continuum_bcm}
\begin{tabular}{cccc}
\hline\hline
Parameter &  Quiescence & Flare\\
\hline
&BMC&\\
$N_{\rm H,1}$ (10$^{22}$ cm$^{-2}$) & 18.9$^{+0.2}_{-0.2}$ & 19.36$^{+0.19}_{-0.19}$\\
$\epsilon$ & 0.995$^{+0.001}_{-0.001}$&0.994$^{+0.001}_{-0.001}$\\
$N_{\rm H,2}$ (10$^{22}$ cm$^{-2}$)&0.30$^{+0.30}_{-0.20}$&2.53$^{+0.36}_{-0.30}$\\
Norm &0.0142$^{+0.0001}_{-0.0001}$&0.0627$^{+0.0006}_{-0.0006}$\\
$kT_{\rm col}$ (keV) &1.43$^{+0.01}_{-0.01}$&0.58$^{+0.01}_{-0.01}$\\
$\alpha$ & 1.25$^{+0.04}_{-0.03}$& 0.186$^{+0.003}_{-0.003}$\\
$f$ &10 (fixed)&10 (fixed)\\
Flux\tnote{a} & $8.00^{+0.05}_{-0.05}$& $45.00^{+0.27}_{-0.27}$\\
&&\\
$\chi^{2}_{\rm r}$(d.o.f.) &  0.96(443)    & 0.98(467) \\
\hline\hline
\end{tabular}
\begin{tablenotes}
\item[a]{Unabsorbed $1.5-20$ \AA\ flux, $\times 10^{-10}$ erg s$^{-1}$ cm$^{-2}$}

\end{tablenotes}
\end{threeparttable}
\end{center}
}
\end{table}

\subsection{Fe lines}

\begin{figure*}
\includegraphics[angle=0,width=1\columnwidth]{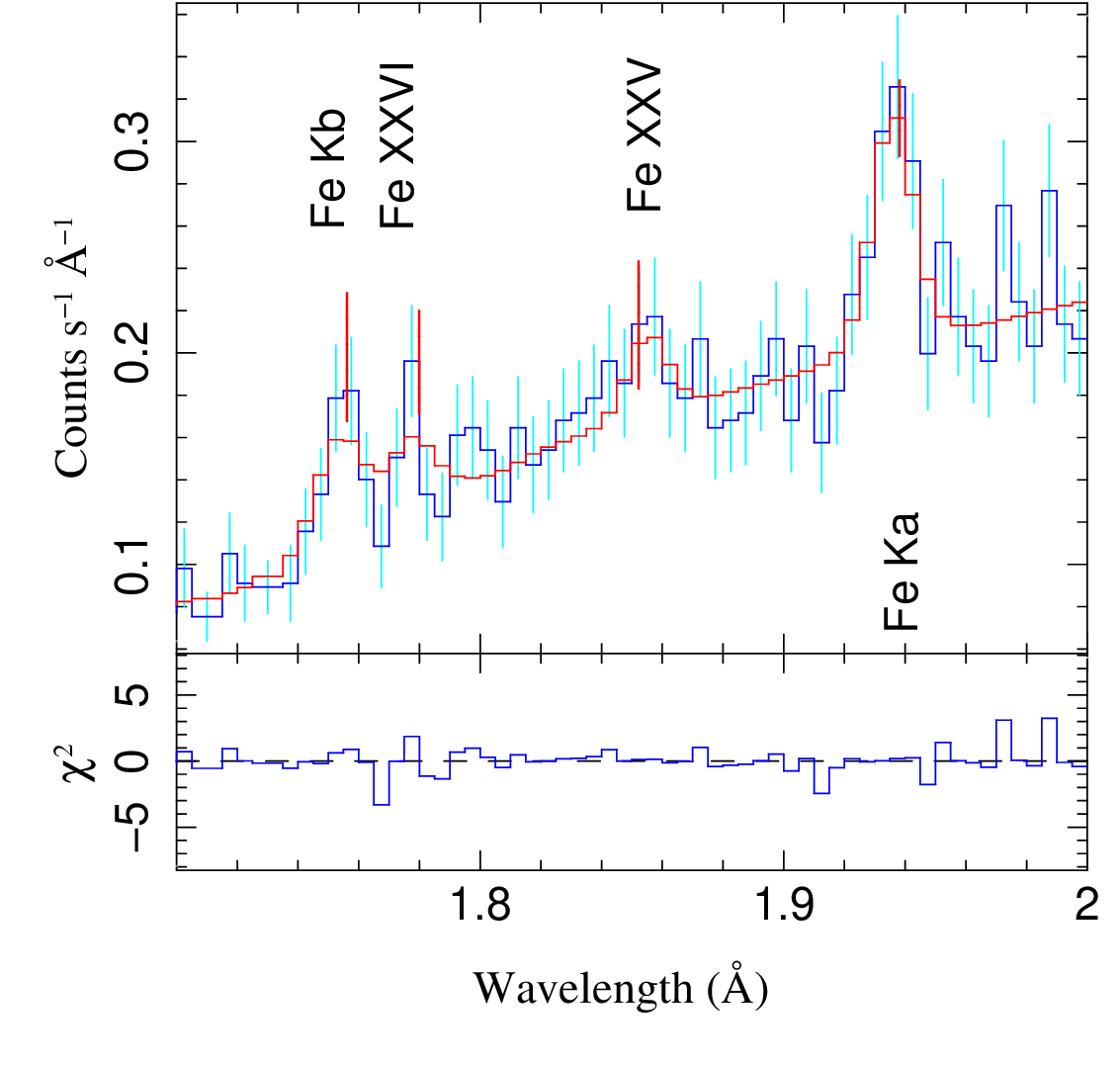}
\includegraphics[angle=0,width=1\columnwidth]{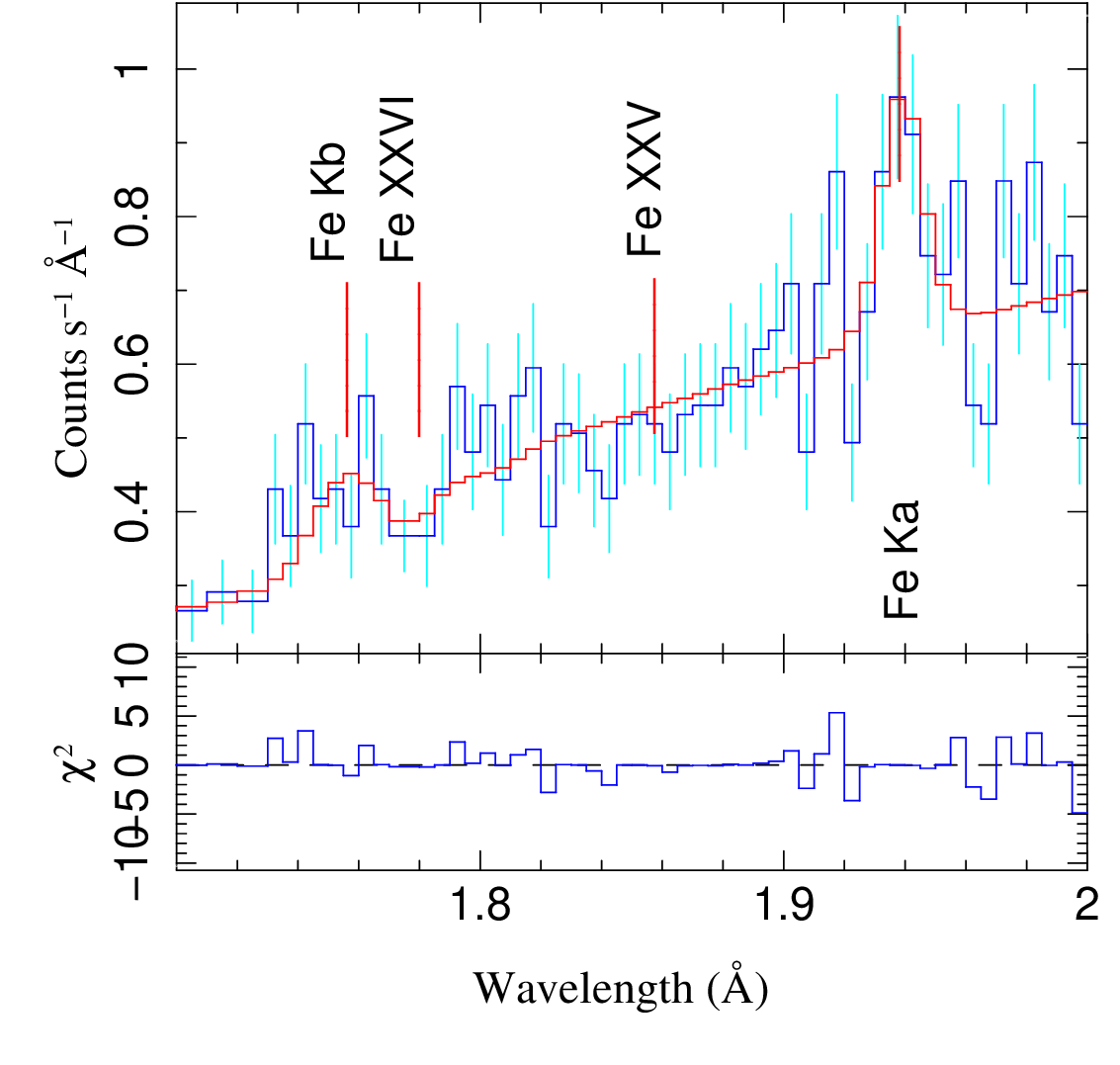}
\caption{\chandra HETG spectra and the best fit model, in 
the 1.6 - 2.1 \AA\ range, during quiescence (left) and flare (right) 
respectively. The lower panels show the quality of fit.}
\label{fig:chandra_spec}
\end{figure*}

The plasma cooling during the flare due to enhanced Compton efficiency is
strongly supported by the analysis of the highly ionised iron lines. 
Figure\,\ref{fig:chandra_spec} shows the Fe K$\alpha$ line region
during quiescence and flare. The quiescence spectra (left panel) show lines from 
the low ionised Fe (K$\alpha$ and K$\beta$ fluorescence, along with 
the Fe K edge at 1.7 keV) as well as from highly ionised states
(Fe \textsc{xxv} He like and Fe \textsc{xxvi} H-like Ly $\alpha$). These 
transitions arise in the circum-source material illuminated by the powerful 
source of X-rays from the accretion onto the compact object. In order to 
measure line fluxes we fitted them as gaussians. Table \ref{tab:lines} lists 
corresponding best fit parameters.

\begin{table*}
{\def\arraystretch{1.3}
\centering
\caption{Fe lines parameters. Numbers without errors have been fixed at the quoted values. } 
\begin{threeparttable}
\label{tab:lines}
\begin{tabular}{cccccccccc}
\hline
& &Quiescence& & & & &Flare & &\\
\hline
Ion  & $\lambda$ & Flux & $\sigma$ & $EW$ & & $\lambda$ & Flux & $\sigma$
& $EW$\\
& & $\times 10^{-6}$ & & &  & & $\times 10^{-6}$ & &  \\
& (\AA) &  (ph s$^{-1}$ cm$^{-2}$) &(\AA) & (m\AA) & & (\AA) &(ph s$^{-1}$
cm$^{-2}$) & (\AA)& (m\AA)  \\
\hline
Fe \ K$\beta$ &1.753$^{+0.003}_{-0.004}$ &190$^{+180}_{-90}$ &0.005
& 21$^{+6}_{-6}$  &  &1.756$^{+0.052}_{-0.052}$  &251$^{+400}_{-150}$ &0.005 & 8$^{+1}_{-7}$\\
Fe \textsc{xxvi} Ly$\alpha$ & 1.777$^{+0.002}_{-0.006}$
&125$^{+80}_{-125}$& 0.005& 7$^{+24}_{-4}$  &   &  1.777$^{+0.002}_{-0.011}$
&-240$^{+290}_{-430}$ &0.005 &  -8$^{+1}_{-13}$  \\
Fe \textsc{xxv} &1.855$^{+0.004}_{-0.000}$ &170$^{+100}_{-50}$ &
0.005 &   18$^{+18}_{-1}$ &   & 1.855$^{+0.004}_{-0.004}$& 0.00
$^{+192}_{-0.01}$&0.005 & 0.0$^{+0.4}_{-0.1}$\\
Fe \ K$\alpha$ & 1.935$^{+0.003}_{-0.003}$ &330 $^{+120}_{-90}$ &
0.005 &  33$^{+11}_{-8}$  &  & 1.939 $^{+0.004}_{-0.003}$ &970
$^{+30}_{-380}$& 0.005 &   31.3$^{+0.7}_{-12.2}$\\
\hline
\end{tabular}
\end{threeparttable}
}
\end{table*}

During the flare the K$\alpha$ fluorescence line from nearly 
neutral Fe remain largely unaffected. The line flux increase in response to 
the increased illumination (higher X-ray continuum) but the equivalent width
($EW$) stays constant (see Table \ref{tab:lines}). In contrast, the highly 
ionised species (He like Fe \textsc{xxv} and H like Fe \textsc{xxvi} Ly 
$\alpha$), disappear. These lines are produced in very high temperature plasma. 
During the episode of enhanced cooling, the Fe ionisation drops and these lines 
should become less prominent. This is indeed confirmed by observations. On the 
other hand, the transition of H like Fe \textsc{xxvi} Ly $\alpha$ seems to
appear in absorption during the flare, hinting to the presence of a warm 
absorber, although the large associated errors prevent a firm conclusion. At 
any rate, the highly ionised Fe lines decrease significantly (or vanish) during 
the flare. 

\section{Discussion}
\label{sec:disc}

The phenomenology presented above can be readily explained as an
episode of Compton plasma cooling during a flare. This behaviour is
predicted in accreting NS systems with moderate X-ray luminosities, undergoing
quasi spherical subsonic accretion \citep{2012MNRAS.420..216S}. 

At a distance of $d\simeq 2$ kpc \citep{2001A&A...370..170A, 2009A&A...507..833M}, the
(un absorbed) luminosity of \so, is $L_{\rm X}\approx 4\times 10^{35}$ erg
s$^{-1}$ during quiescence. At such luminosities, direct (Bondi)
wind accretion is hampered by the need of the plasma to cool. If
$t_{\rm cool}>> t_{\rm freefall}$ the matter undergoes subsonic
settling accretion and a quasi spherical shell appears around the
NS between the magnetospheric radius and the Bondi radius. 
The presence of a convective shell is supported by the
presence of mHz QPOs in the light curve. These QPOs might reflect  
the convective motions inside the shell.

On the other hand, the luminosity of 4U1700$-$37 in quiescence is very close 
to the threshold predicted for a phase transition from radiative to Compton 
cooling, $\sim 3\times 10^{35}$ erg s$^{-1}$ \citep{2013MNRAS.428..670S}. 
Hence, \so\ is very prone to such transitions which may easily be caused by 
a local perturbation in the accretion flow. Following the transition, the 
enhanced efficiency of the Compton cooling further increases the ability of matter 
to enter the magnetosphere and accrete. This runaway process leads to a flare. 
During the observed flare, the luminosity reached  $L_{\rm X}\approx 2\times 
10^{36}$\,erg\,s$^{-1}$, still too low for the direct Bondi accretion regime. 

In our observation, the flare is preceded by an off-state (see 
Fig.\,\ref{fig:chandra_lc}). As a possible explanation for these 
observational facts, one may consider a clumped stellar wind, where 
the medium between clumps is strongly rarefied \citep{2012MNRAS.421.2820O}.    
In this case, a perturbation caused by the ingestion of a wind clump (flare) 
would be preceded by a short period of time when the wind in the vicinity of 
the NS is void (off-state).   

At the quasi-spherical settling accretion stage, the neutron star
equilibrium spin period can be very long, about thousand seconds for
the canonical neutron star magnetic field $10^{12}$~G and typical
stellar wind velocities of about 1000 km~s$^{-1}$
\citep{2012MNRAS.420..216S}. The spin period is almost directly
proportional to the NS magnetic field, and for a highly magnetised NS
can be a few ten thousand seconds or even longer \citep[36.2 ks in AX
J1910.7$+$091]{2017MNRAS.469.3056S}. This may explain the
non-detection of coherent pulsations. 

Despite the lack of pulse detection, other pieces of
evidence support the presence of a magnetised NS. The temperature of
the soft photons is rather high during
quiescence. To reconcile the high colour temperature of the soft
emitting region $T_{\rm col}\sim 1.4$ keV with the low intrinsic
luminosity ( $L_{\rm X}\sim 10^{35}$ erg s$^{-1}$) a small
emission area must be invoked. Assuming that the source is radiating
as a black body of area $4\pi R_{W}^{2}$, the radius of the emitting area
would be $R_{W}=0.3\sqrt{L_{34}} (kT_{col}/1 \rm
kev)^{-2}\simeq 0.9$ km, which is only compatible with a hot spot over the NS
surface. On the other hand, during the flare this radius turns out to
be much larger, of the order of 12\,km, comparable in size to the entire
NS. 

\section{Conclusions}
\label{sec:conc}

We present empirical evidence of plasma Compton cooling during a flare
in 4U1700$-$37. This is supported by the analysis of the X-ray continuum as
well as the disappearance of the highly ionised Fe lines. This behaviour can be
explained by the sudden accretion of the hot shell that forms around
the NS when a transition from a radiative cooling regime to a much more
efficient Compton cooling, occurs. The predicted luminosity for such
transitions, namely $\sim 3\times 10^{35}$ erg s$^{-1}$
\citep{2013MNRAS.428..670S} is very close to where \so stays during
quiescence. The presence of such hot shell is further supported by the
detection of mHz QPOs produced by convection cells in the shell. To
reconcile the high plasma temperature with the low $L_{X}$ a small
emitting area $R_{W}\sim 1$ km, must be invoked, only compatible with
a hot spot on a NS surface. Therefore, a magnetised NS is strongly
favoured by the available data. The lack of coherent pulsations may indicate 
a very long spin period of a strongly magnetised neutron star with 
$B>10^{13}$~G.

\section*{Acknowledgements}

This research has been supported by the grant ESP2014-53672-C3-3P. AB
acknowledges support from STScI award 44A-1096046. JJRR acknowledges
support from MECD fellowship PRX17/00114. This research has made use 
of the \textsc{isis} functions provided by 
ECAP/Remeis observatory and MIT. 
We
thank \chandra director for the approval of the Director's
Discretionary Time observation and the anonymous referee
whose comments improved the content of the paper. 




\bibliographystyle{mnras}
\bibliography{bibliografia} 







\bsp	
\label{lastpage}
\end{document}